\documentclass[10pt, conference, letterpaper]{IEEEtran}
\usepackage[utf8]{inputenc}
\usepackage{graphicx}
\usepackage{amsmath}
\usepackage{amsfonts}
\usepackage{amssymb}
\usepackage{booktabs}
\usepackage{tabularx}
\usepackage{tikz}
\usepackage{cite}
\usepackage{url}
\usepackage{float}
\usepackage[caption=false,font=footnotesize]{subfig}
\usepackage{xcolor}
\usepackage{lipsum}

\usetikzlibrary{shapes,arrows,positioning}
\newcommand{\keywords}[1]{\par\noindent\textbf{Keywords:} #1}

\begin{document}

\title{Analyzing Crime Discourse in U.S. Metropolitan Communities on Reddit: Trends, Influences, and Insights}

\author{\IEEEauthorblockN{Deepit Sapru}
\IEEEauthorblockA{University of Illinois Urbana-Champaign \\
dsapru2@illinois.edu}
}

\maketitle

\begin{abstract}
The relationship between crime and the media has long been a focal point of academic research, with traditional media playing a significant role in shaping public perceptions of safety and community well-being. However, the advent of social media has introduced a new dimension to this discourse, offering unique platforms for user-driven discussions. Despite the prominence of social media, research examining its impact on crime-related discourse remains limited. This paper investigates crime-related discussions across Reddit communities representing 384 Metropolitan Areas in the United States. By analyzing user submissions, we identify key trends in crime discourse, including the higher prevalence of such discussions in larger metropolitan areas and communities with more liberal political leanings. Interestingly, we find that reported crime rates do not strongly influence the frequency or intensity of these discussions. These findings provide novel insights into how social media platforms, like Reddit, shape public narratives around crime, highlighting the growing need to examine digital spaces as influential mediators of public perception.
\end{abstract}

\section{Introduction}

The intersection of crime and media representation has constituted a significant area of scholarly inquiry for decades, with traditional media outlets extensively studied for their role in shaping public understanding of crime and safety \cite{heath1996mass}. Television news, newspapers, and radio broadcasts have historically functioned as primary sources of crime information, often influencing community perceptions and policy decisions \cite{sacco1995media}. However, the digital revolution has fundamentally transformed media consumption patterns, with social media platforms increasingly supplanting traditional media as primary information sources \cite{nielsen2016happening}.

Reddit, as a prominent social media platform, offers a unique environment for examining crime discourse due to its community-driven structure and geographical subreddits. Unlike traditional media with centralized editorial control, Reddit enables user-generated content that may more accurately reflect grassroots concerns and discussions \cite{jhaver2019transparency}. This paper addresses a critical gap in the literature by systematically analyzing crime-related discussions across 384 Metropolitan Statistical Areas (MSAs) in the United States, examining how these discussions correlate with demographic, political, and crime statistics.

Our research employs a comprehensive data collection methodology, gathering 161,166 posts from geographically-defined subreddits over a two-month period. Through textual analysis and statistical examination, we investigate whether crime discourse on social media reflects actual crime rates or is influenced by other socio-political factors. This approach builds upon previous work examining crime discourse on platforms like Twitter \cite{curiel2020crime} while addressing the unique community structure of Reddit.

The significance of this research extends beyond academic interest, as understanding crime discourse patterns on social media has implications for public policy, community engagement, and media literacy. As social media platforms increasingly shape public opinion \cite{walker2021news}, comprehending the dynamics of crime-related discussions becomes essential for addressing potential misinformation and understanding community concerns.

This paper makes several key contributions: First, we develop a robust methodology for identifying crime-related content across diverse online communities. Second, we provide empirical evidence regarding the relationships between crime discourse, population characteristics, political leanings, and actual crime statistics. Third, we offer insights into the behavioral patterns of users who frequently engage in crime-related discussions. Finally, we contextualize our findings within broader discussions about the role of social media in shaping public perception of crime and safety.

\section{Background and Literature Review}

\subsection{Traditional Media and Crime Representation}

Traditional media's relationship with crime reporting has been extensively documented in academic literature. Research consistently demonstrates that media coverage often exaggerates crime prevalence and severity, contributing to what scholars term "crime fear" or "mean world syndrome" \cite{shanahan1999television}. Heath and Gilbert \cite{heath1996mass} found that increased television news consumption correlated with heightened distrust in local communities and broader society, regardless of actual crime statistics.

The political implications of crime reporting have also received significant attention. Liu et al. \cite{liu2021campaign} demonstrated how crime-focused campaign advertisements during the 2016 U.S. presidential election influenced voting patterns, with viewers who consumed more crime-related content tending to support conservative candidates. This relationship highlights how crime discourse can transcend mere public safety concerns and become embedded in political narratives.

Institutional responses to crime perception have similarly been shaped by media representations. The implementation of policies like "stop and frisk" and the theoretical underpinnings of "Broken Windows" policing \cite{wilson1982broken} emerged from particular understandings of crime and community safety that were often amplified through media channels. These policies, while controversial, demonstrate how media-shaped perceptions can translate into concrete social interventions \cite{meares2014law}.

\subsection{Social Media Platforms and Crime Discourse}

The transition from traditional to social media as primary news sources has introduced new dynamics in crime discourse. Walker and Masta \cite{walker2021news} document that nearly half of Americans now consume news through social media platforms, representing a fundamental shift in information consumption patterns. This transition has occurred alongside the decline of traditional print and broadcast media \cite{nielsen2016happening}.

Research on crime discourse specific to social media platforms remains relatively limited compared to traditional media studies. Curiel et al. \cite{curiel2020crime} examined crime-related tweets in Latin American contexts, finding that Twitter discussions did not correlate strongly with documented crime rates but effectively captured community anxiety about crime. This discrepancy between discussion frequency and actual crime statistics suggests social media may amplify perceived rather than actual crime threats.

Social media platforms also present unique challenges regarding misinformation and content moderation \cite{phadke2020many}. Unlike traditional media with established editorial standards, social media platforms often struggle with balancing free expression and preventing the spread of false information \cite{zannettou2021won}. This challenge is particularly relevant for crime discourse, where misinformation can have real-world consequences, as demonstrated during events like the Boston Marathon bombing \cite{potts2013interfaces}.

\subsection{Reddit as a Research Platform}

Reddit's community-driven structure offers distinctive opportunities for analyzing geographically-based discussions. The platform's subreddit system allows for the examination of location-specific conversations while maintaining the scale necessary for robust statistical analysis \cite{rajadesingan2021political}. Previous research has examined various aspects of Reddit communities, including political discourse \cite{rajadesingan2021political}, misinformation patterns \cite{ribeiro2021platform}, and community migration behaviors \cite{davies2021multi}.

The geographical nature of many Reddit communities, while not enforced through technical means, often results in localized discussions \cite{scellato2010distance}. Blackburn et al. \cite{blackburn2014cheating} found that even on platforms without explicit geographical features, users tend to form communities with local characteristics. This tendency makes Reddit an valuable platform for examining how crime discourse varies across different metropolitan contexts.

Content moderation on Reddit represents another important consideration. Unlike centrally-moderated traditional media, Reddit communities are typically moderated by volunteers with varying approaches and standards \cite{jhaver2019transparency}. This decentralized moderation can lead to significant variation in discussion quality and focus across different geographical subreddits, potentially influencing crime discourse patterns.

\section{Methodology and Data Collection}

\subsection{Data Acquisition Framework}

Our research employed a comprehensive data collection strategy focused on Reddit communities representing the 384 Metropolitan Statistical Areas (MSAs) in the United States as defined by the U.S. Census Bureau \cite{census2020metropolitan}. We developed a specialized crawler utilizing Reddit's REST API \cite{reddit2022api} to gather posts from each geographical subreddit between April 5, 2022, and May 31, 2022. The crawler operated on an hourly schedule, ensuring comprehensive coverage while minimizing API rate limit issues.

The data collection infrastructure employed a dual-database approach for robustness and analytical flexibility. A CockroachDB instance served as the system of record, storing raw post data in JSON format, while an ElasticSearch database facilitated efficient text analysis and keyword extraction. This architecture allowed for both reliable data preservation and performant analytical processing. The entire system operated within Docker containers managed by Kubernetes, ensuring high availability and automatic recovery from potential failures.

During the two-month collection period, we gathered 161,166 posts across all monitored subreddits. The distribution of posts followed expected patterns relative to population size, though with notable variations potentially attributable to differing moderation policies and community engagement levels across subreddits. This extensive dataset provided a robust foundation for analyzing crime discourse patterns across diverse metropolitan contexts.

\subsection{Crime Discourse Identification}

Identifying crime-related content required developing a precise methodology for classifying posts based on their textual content. We compiled a comprehensive list of 124 crime-related keywords derived from FBI Uniform Crime Reporting categories \cite{fbi2020ucr}, focusing particularly on violent crimes and their common synonyms. The keyword list included terms such as "assault," "burglary," "shooting," "robbery," and related variants, ensuring broad coverage of crime-related discussions while maintaining specificity.

We implemented a regular expression-based classification system that analyzed both post titles and linked URLs for crime-related keywords. Posts containing at least one keyword from our predefined list were flagged as crime-related. To validate this approach, we conducted manual analysis of a 400-post sample, achieving 91\% classification accuracy. This accuracy rate compares favorably to similar methodologies employed in social media research \cite{curiel2020crime}, though we acknowledge potential limitations regarding false negatives due to dog-whistles or localized terminology.

The classification process identified 6,792 crime-related posts from the total dataset, representing approximately 4.2\% of all collected content. This proportion varied significantly across different metropolitan areas, with some communities exhibiting crime discussion rates exceeding 15\% while others showed minimal crime-related content. These variations formed the basis for our subsequent analysis of factors influencing crime discourse prevalence.

\subsection{Contextual Data Integration}

To contextualize Reddit discourse patterns, we integrated multiple external datasets characterizing each metropolitan area. Population data derived from U.S. Census Bureau statistics \cite{census2020age} enabled normalization of discussion volumes and examination of size-related patterns. We categorized MSAs into four population classes: large (over 1,000,000), medium (500,000-1,000,000), small (200,000-500,000), and very small (under 200,000), facilitating comparative analysis across differently-sized communities.

Political context incorporated 2020 presidential election results \cite{ap2020election}, specifically the percentage of voters supporting Democratic candidates within each MSA. We established five political categories based on voting patterns: very conservative (under 37.5\% Democratic), conservative (37.5-47.5\%), moderate (47.5-52.5\%), liberal (52.5-62.5\%), and very liberal (over 62.5\% Democratic). This granular categorization allowed for detailed examination of political correlations with crime discourse.

Documented crime rates came from the FBI's Uniform Crime Reporting system \cite{fbi2020ucr}, focusing specifically on violent crime rates per 100,000 residents. We selected violent crime statistics due to their more consistent reporting standards across jurisdictions compared to property crimes. All crime rate data underwent min-max normalization to enable comparative analysis with similarly normalized crime discussion metrics.

\begin{table}[htbp]
\caption{Metropolitan Statistical Areas with Highest Crime Discussion Rates}
\label{tab:high_crime_discussion}
\centering
\begin{tabular}{p{2cm}cc}
\toprule
\textbf{Metropolitan Statistical Area} & \textbf{\% Crime Posts} & \textbf{Crime Posts} \\
\midrule
Seattle-Tacoma-Bellevue, WA & 18.2 & 344 \\
Anniston-Oxford, AL & 15.7 & 8 \\
Lewiston-Auburn, ME & 15.4 & 14 \\
Bay City, MI & 14.0 & 7 \\
Lewiston, ID-WA & 13.0 & 7 \\
Los Angeles-Long Beach-Anaheim, CA & 12.0 & 257 \\
Portland-Vancouver-Hillsboro, OR-WA & 11.4 & 287 \\
Rocky Mount, NC & 11.3 & 7 \\
New York-Newark-Jersey City, NY-NJ-PA & 10.7 & 196 \\
Flint, MI & 9.7 & 10 \\
\bottomrule
\end{tabular}
\end{table}

\section{Empirical Analysis and Findings}

\subsection{Population Characteristics and Crime Discourse}

Our analysis revealed significant relationships between metropolitan area population size and crime discussion prevalence. Large metropolitan areas (population over 1,000,000) exhibited substantially higher normalized crime discussion scores (0.260) compared to medium (0.166), small (0.173), and very small (0.196) areas. This pattern suggests that population density and urban scale may amplify crime-related discussions independent of actual crime statistics.

The distribution of crime discourse across population categories followed distinct patterns. While very small metropolitan areas included both communities with exceptionally high crime discussion rates and communities with no crime-related posts, large metropolitan areas consistently maintained moderate to high levels of crime discourse. This consistency indicates that population size establishes a baseline level of crime discussion, though other factors influence specific discussion volumes.

Notably, the relationship between population size and crime discussion intensity was not linear. Medium-sized metropolitan areas actually showed slightly lower average crime discussion than small areas, suggesting potential threshold effects where discussion patterns shift at specific population levels. This nonlinear relationship underscores the complexity of factors influencing online crime discourse and cautions against simplistic population-based predictions.

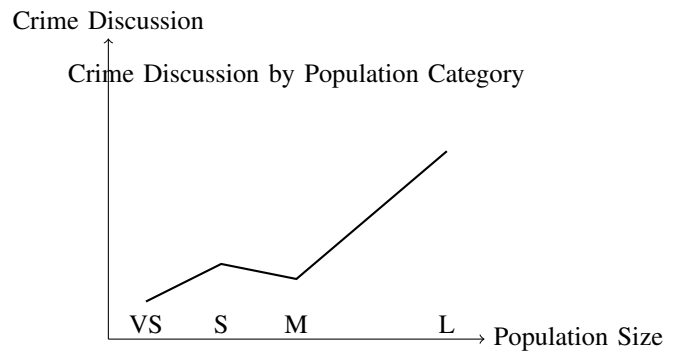
\begin{figure}[htbp]
\centering
\begin{tikzpicture}
\draw[->] (0,0) -- (5,0) node[right] {Population Size};
\draw[->] (0,0) -- (0,4) node[above] {Crime Discussion};
\draw[thick] (0.5,0.5) -- (1.5,1) -- (2.5,0.8) -- (4.5,2.5);
\node at (0.5,0.2) {VS};
\node at (1.5,0.2) {S};
\node at (2.5,0.2) {M};
\node at (4.5,0.2) {L};
\node at (2.5,3.5) {Crime Discussion by Population Category};
\end{tikzpicture}
\caption{Relationship between population size categories and crime discussion intensity. VS=Very Small, S=Small, M=Medium, L=Large metropolitan areas.}
\label{fig:population_discussion}
\end{figure}

\subsection{Political Correlations with Crime Discourse}

Political characteristics emerged as strong predictors of crime discussion prevalence across metropolitan areas. Our analysis demonstrated a clear positive relationship between liberal voting patterns and crime discourse intensity. Very liberal metropolitan areas exhibited the highest average normalized crime discussion scores (0.274), followed by moderate (0.207), liberal (0.201), conservative (0.179), and very conservative (0.170) areas.

This political gradient remained evident even when controlling for population size. Among large and medium metropolitan areas, those with higher proportions of Democratic voters consistently showed elevated crime discussion rates. The most conservative large metropolitan area in our dataset, Provo-Orem, Utah, recorded no crime-related posts during the study period, while strongly liberal areas like Seattle, Portland, and New York City exhibited crime discussion rates exceeding 10\%.

The divergence between crime discussion and documented crime rates across political categories was particularly noteworthy. Very liberal metropolitan areas, while showing the highest crime discussion levels, actually had lower average documented violent crime rates than conservative areas. This discrepancy suggests that crime discourse on social media reflects factors beyond objective crime statistics, potentially including political identity, media consumption patterns, or community values regarding public safety discussions.

\subsection{Documented Crime Rates and Online Discourse}

Contrary to intuitive expectations, we found no strong correlation between documented violent crime rates and the prevalence of crime-related discussions on Reddit. Metropolitan areas with the highest violent crime rates, including Birmingham-Hoover, Alabama (1,397.2 per 100,000) and Farmington, New Mexico (1,199.8 per 100,000), showed only moderate crime discussion rates of 2.52\% and 1.61\% respectively.

The disconnect between actual crime statistics and online discussion became particularly evident when examining normalized difference scores between crime discussion and documented crime rates. Metropolitan areas with the largest positive differences (indicating disproportionately high crime discussion relative to actual crime) were predominantly large, liberal areas in coastal regions, while areas with large negative differences (high crime but low discussion) tended to be smaller, conservative communities in the South and Midwest.

This misalignment between objective crime indicators and subjective online discussions has important implications for public understanding of crime patterns. If social media discourse disproportionately represents certain types of communities or perspectives, it may distort public perception of actual crime risks and appropriate policy responses. This potential distortion underscores the importance of critical media literacy regarding crime-related social media content.

\begin{table}[htbp]
\caption{Metropolitan Areas with Highest Documented Violent Crime Rates}
\label{tab:high_crime_rates}
\centering
\begin{tabular}{lcc}
\toprule
\textbf{Metropolitan Statistical Area} & \textbf{Crime Rate} & \textbf{\% Crime Discussion} \\
\midrule
Birmingham-Hoover, AL & 1,397.2 & 2.52 \\
Farmington, NM & 1,199.8 & 1.61 \\
Anchorage, AK & 1,194.6 & 2.36 \\
Kansas City, MO-KS & 1,159.4 & 3.16 \\
Memphis, TN-MS-AR & 1,120.5 & 3.43 \\
Winston-Salem, NC & 1,073.6 & 2.32 \\
Albuquerque, NM & 1,043.4 & 4.53 \\
Danville, IL & 935.7 & 0.00 \\
Pine Bluff, AR & 895.4 & 12.50 \\
Odessa, TX & 881.8 & 4.30 \\
\bottomrule
\end{tabular}
\end{table}

\section{User Behavior and Sentiment Analysis}

\subsection{Sentiment Patterns in Crime Discourse}

Our analysis of user sentiment revealed distinct differences between individuals who frequently post crime-related content and those who do not. Focusing on the r/nyc subreddit as a case study, we examined 1,197 unique posters, identifying 134 users who had posted about crime at least once. Sentiment analysis of these users' comments and posts, both within and outside the New York City community, showed consistently more negative sentiment among crime posters.

Crime-posting users exhibited average sentiment polarity scores of -0.07 both within r/nyc and in their broader Reddit activity. In contrast, non-crime posters showed positive average sentiment scores of 0.09 within the community and 0.10 elsewhere on the platform. This sentiment gap suggests that users who frequently engage with crime content may have generally more negative outlooks that extend beyond specific crime-related discussions.

The consistency of sentiment patterns across different subreddits for crime-posting users indicates that negative expression may represent a stable characteristic of these individuals' online behavior rather than a situation-specific response to local crime conditions. This finding aligns with research suggesting that certain personality traits or worldviews may predispose individuals to focus on negative societal aspects \cite{medhat2014sentiment}.

\subsection{Community Engagement Patterns}

Analysis of user engagement across different subreddits revealed distinctive patterns between crime-posting and non-crime-posting users. Using the Jaccard Index to measure community overlap \cite{fletcher2018comparing}, we found that crime posters showed strong connections to specific crime-focused communities like CrimeInNYC and NYStateOfMind, with overlap scores of 0.83 and 0.12 respectively.

Non-crime posters demonstrated stronger connections to location-specific hobby and interest communities, including nycrail (0.81), RunNYC (0.48), and various neighborhood subreddits. This pattern suggests that while crime posters maintain some local engagement, their community connections are more diffuse than those of non-crime posters, who show tighter integration with geographically-based interest communities.

The political subreddit engagement patterns further distinguished these user groups. Crime posters showed higher engagement with conservative-leaning communities like LockdownSkepticism and Conservative, while non-crime posters more frequently participated in moderate and liberal political spaces like neoliberal and VoteDEM. This political divergence occurred despite New York City's overall liberal orientation, suggesting that crime posters may represent a politically distinct subgroup within their communities.

\begin{figure}[htbp]
\centering
\begin{tikzpicture}
\draw (0,0) circle (1.5cm) node {Crime Posters};
\draw (3,0) circle (1.5cm) node {Non-Crime Posters};
\node at (1.5,0) {};
\draw (0,0) -- (1,1);
\draw (3,0) -- (2,1);
\node at (0,1.8) {Conservative};
\node at (0,1.5) {Communities};
\node at (3,1.8) {Local Hobby};
\node at (3,1.5) {Communities};
\node at (1.5,-2) {Distinct Community Engagement Patterns};
\end{tikzpicture}
\caption{Visualization of distinct community engagement patterns between crime-posting and non-crime-posting users}
\label{fig:engagement_patterns}
\end{figure}

\section{Discussion and Implications}

\subsection{Interpreting the Crime Discourse Landscape}

Our findings present a complex picture of crime discourse on social media, challenging several common assumptions about the relationship between online discussions and real-world conditions. The strong correlation between liberal political leanings and crime discussion prevalence contradicts narratives that conservative communities focus more heavily on crime issues. Instead, our results suggest that crime discourse may serve different social or political functions in different community contexts.

The disconnect between documented crime rates and discussion frequency raises important questions about what drives online crime conversations. If these discussions do not primarily reflect objective crime conditions, they may instead express broader community anxieties, political positions, or responses to media narratives. This interpretation aligns with research suggesting that social media often amplifies perceived risks rather than actual threats \cite{curiel2020crime}.

The demographic patterns we observed—specifically the elevated crime discussion in larger metropolitan areas—may reflect genuine differences in urban experience, including higher population density, greater anonymity, and more frequent stranger interactions. However, the magnitude of the difference suggests that additional factors, such as media ecosystem differences or community norms around public discussion, also play significant roles.

\subsection{Practical Implications and Applications}

Our research has several practical implications for community leaders, law enforcement agencies, and media organizations. First, the demonstrated misalignment between online crime discourse and actual crime statistics highlights the potential for social media to distort public understanding of safety issues. Organizations communicating about crime should recognize this potential distortion and consider strategies to provide context and accurate statistics.

Second, the political dimensions of crime discourse suggest that crime discussions on social media may function partially as expressions of political identity rather than straightforward reporting of community conditions. Recognizing this dimension could help community leaders interpret crime-related social media activity more accurately and respond appropriately to constituent concerns.

Third, the behavioral patterns of frequent crime posters—including their more negative general sentiment and distinct community affiliations—suggest that crime discourse may be driven by a subset of community members with particular characteristics. Understanding these patterns could inform more effective community engagement strategies and help identify potential misinformation sources.

\subsection{Limitations and Future Research Directions}

While our study provides comprehensive analysis of crime discourse patterns, several limitations warrant acknowledgment. Our keyword-based classification approach, while validated through manual review, may miss nuanced crime discussions or local terminology. Future research could incorporate more advanced natural language processing techniques to capture broader semantic patterns.

The temporal scope of our data collection—approximately two months—provides a snapshot of discussion patterns but may not capture seasonal variations or responses to specific events. Longitudinal studies tracking crime discourse over extended periods could reveal how these patterns evolve in response to real-world developments and media cycles.

Our focus on Reddit, while providing valuable insights into one prominent platform, limits generalizability to other social media environments with different user demographics and communication norms. Comparative studies across multiple platforms—such as Facebook, Nextdoor, and Twitter—could illuminate platform-specific dynamics in crime discourse.

Future research could also explore more nuanced aspects of crime discourse, including the types of crimes most frequently discussed, the framing of these discussions, and their relationship to local media coverage. Additionally, investigating intervention strategies to ensure accurate crime perception in social media-saturated environments represents an important direction for applied research.

\section{Conclusion}

This research provides systematic examination of crime-related discussions across U.S. metropolitan areas on Reddit, revealing distinct patterns that challenge simplistic assumptions about crime discourse. Our findings demonstrate that crime discussion prevalence correlates strongly with population size and political leanings but shows little relationship to documented crime rates. These patterns suggest that social media crime discourse serves complex social functions beyond simple information sharing about local safety conditions.

The behavioral characteristics of frequent crime posters—including generally negative sentiment and distinct community affiliations—further complicate the interpretation of crime-related social media content. Rather than representing broad community concerns, crime discourse may often reflect the perspectives and motivations of specific subgroups within communities.

As social media platforms continue to evolve as primary spaces for public discourse, understanding the dynamics of crime-related discussions becomes increasingly important for accurate public perception and effective policy responses. Our research contributes to this understanding by documenting systematic patterns in how crime is discussed across different community contexts and identifying factors that shape these discussion patterns.

Future research should build upon these findings to develop more sophisticated models of crime discourse dynamics, examine cross-platform patterns, and explore strategies for ensuring that online discussions accurately reflect community safety conditions. Through continued investigation of these issues, researchers can help ensure that social media serves as a constructive rather than distorting influence on public understanding of crime and safety.

\bibliographystyle{IEEEtran}
\bibliography{references}

\end{document}